\documentclass[12pt]{article}
\usepackage{graphicx}% Include figure files
\newcommand{\sss}{\scriptscriptstyle}

\newcommand {\be}{\begin{equation}} % start equation
\newcommand{\ee}{\end{equation}}    % end equation

\def\dds1{\frac{\partial}{\partial s_1}}

\def\d{d\kern-0.8 ex\vrule height 1.3 ex depth-1.24 ex width 0.7 ex
\kern 0.15 ex}
\def\D{D\kern-1.7 ex\vrule height .87 ex depth-0.8 ex width 0.7 ex
\kern 0.95 ex}

\newcommand{\nin}{\nu_{in}}
\newcommand{\nen}{\nu_{en}}

\newcommand{\nei}{\nu_{ei}}
\newcommand{\nue}{\nu_{ei} + \nu_{en}}

\begin{document}
\baselineskip 20 pt

\begin{center}

\Large{\bf Energy flux of Alfv\'en waves in weakly ionized plasma }

\end{center}

%\vspace{0.3cm}

\begin{center}

 J. Vranjes$^1$, S. Poedts$^1$, B. P. Pandey$^2$, and B. De Pontieu$^3$

{\em $^1$Center for Plasma Astrophysics, Celestijnenlaan 200B, 3001 Leuven,
 Belgium.}

%\vspace{5mm}

  {\em  $^2$Department of Physics, Macquarie University, Sydney, NSW
2109, Australia.}

%\vspace{5mm}

{\em 
$^3$Lockheed Martin Solar and Astrophysics Lab, 3251 Hanover St., Org. ADBS,
Bldg. 252, Palo Alto, CA 94304 94304, USA.}
\end{center}

\vspace{1cm}

{\bf Abstract:} The overshooting convective motions in the solar photosphere are frequently proposed as the source for the excitation of Alfv\'en
waves. However, the photosphere is a)~very weakly ionized, and, b)~the
dynamics of the plasma particles in this region is heavily influenced by the plasma-neutral collisions. The purpose of this work is to check the consequences of these two facts on the above scenario and their effects on
the electromagnetic waves.
It is shown that the ions and electrons in the photosphere are both un-magnetized; their collision frequency with
neutrals is much larger than the gyro-frequency. This implies that eventual Alfv\'en-type electromagnetic
perturbations must involve the neutrals as well. This has the following serious consequences: i)~in the presence of
perturbations, the whole fluid (plasma + neutrals) moves; ii)~the Alfv\'en velocity includes the total (plasma +
neutrals) density and is thus considerably smaller compared to the collision-less case; iii)~the perturbed velocity
of a unit volume, which now includes both plasma and neutrals, becomes much smaller compared to the ideal
(collision-less) case; and iv)~the corresponding wave energy flux for the given parameters becomes much smaller
compared to the ideal case.

\vfill

\pagebreak

%==============================================================================
\section{Introduction}
%==============================================================================

In a weakly ionized but highly collisional medium, a propagating Alfv\'en wave also involves the motion of the
neutrals that are present in the medium. This is due to the friction between charged particles and neutrals. The
effect has been described in the literature,  first  by  Tanenbaum and Mintzer (1962),   Woods (1962), Jephcott and   Stocker (1962), and in many subsequent works, e.g.,  Kulsrud and Pierce (1969), Pudritz (1990), Haerendel (1992), De Pontieu and Haerendel (1998),  Watts and  Hanna (2004). This fundamental result is valid for any weakly ionized plasma, including the plasma in the lower solar atmosphere.

The Alfv\'en wave has been a very popular tool  in the scenarios and models dealing with the heating of
upper solar atmosphere. A necessary ingredient in such models is an efficient and abundant source for the excitation
of these waves, which acts permanently and generates waves throughout the solar atmosphere. Very frequently it is
assumed that the omnipresent overshooting convective motions in the photosphere could serve for this purpose. The
amount of thermal energy per unit volume in the solar corona is in fact extraordinarily  small in comparison with the
lower (and much colder) layers of the solar atmosphere. This is due to the rapidly decreasing density with altitude.
On the other hand, the complete photosphere is covered by overshooting convective gas motions with typical velocities
of about $0.5\;$km/s, that may go up to $2\;$km/s. The kinetic energy per cubic meter stored in this macroscopic
motion of a mainly neutral gas exceeds for several orders of magnitude the internal energy density in the corona.
Clearly, only a tiny fraction of the convective kinetic energy of the neutral gas would be sufficient to heat the
higher layers to the given temperatures. Such a scenario is attractive in view of the fact that this macroscopic
motion in the lower atmosphere is permanent and widespread throughout the solar surface. However, the photosphere is
very weakly ionized and it is also a strongly collisional mixture of the tiny plasma component and the predominantly
neutral (uncharged) gas.

The energy flux of the Alfv\'en waves is given by $m_i n_i v_i^2 c_a/2$, where $c_a$ is the  Alfv\'en velocity and
$v_i$ is the perturbed velocity of ions involved in the oscillations. Typically, in the estimate of the flux in the
photosphere, this perturbed velocity taken is  of the same order as the macroscopic convective motion mentioned above
(Hollweg 1981).

In the present work, we focus on the physics involved in the propagation of the Alfv\'en wave in a weakly ionized
plasma like the solar photosphere. Using simple and reliable physical arguments and widely accepted plasma theory, we
discuss the flux of the Alfv\'en waves under these circumstances. It will be shown that, if we assume the existence
of the necessary electromagnetic perturbations in such a weakly ionized medium, the energy flux of the waves is in
fact much lower compared to what is usually expected from estimates based on ideal magnetohydrodynamics. This is due to the fact that the photospheric gas dynamics is heavily influenced by collisions. More precisely, in the presence of some accidental
electromagnetic perturbations, which in the first step involve plasma species (electrons and ions) only, the neutral
atoms respond to these electromagnetic perturbations due to the strong friction. This, and the fact that the
ionization ratio is rather small (viz.\ of the order of $10^{-4}$), results in very small amplitudes of the perturbed
velocity of the {\em total} plasma-gas fluid.

\begin{table*}
\caption{Collision frequencies (in Hz) and  magnetization ratio of
electrons and protons  in the photosphere for two altitudes $h$ (in km) and for the magnetic field $B_0=10^{-2}$
T. } \label{table:1} \centering
\begin{tabular}{c c c c c c c  }     % 11 columns
\hline\hline
 h  & $\nu_{in}$ & $\nu_{ii}$ & $\nu_{en}$ & $\nu_{ei}$ &
$\Omega_i/\nu_{it}$ & $\Omega_e/\nu_{et}$ \\
\hline
   0   & $1.6\cdot 10^9$ &$5\cdot 10^7$     & $ 1.3\cdot 10^{10}$ &
$1.5\cdot 10^9$ & $6\cdot 10^{-4}$& $1.1\cdot 10^{-1}$ \\
   250 & $2.6\cdot 10^8$ & $3.8\cdot 10^6$  & $2.2\cdot 10^9$ &
$1.2\cdot 10^8$ & $3.6\cdot 10^{-3}$ & $7.3\cdot 10^{-1} $  \\
 \hline
\end{tabular}
\end{table*}

%==============================================================================
\section{Physics of  weakly ionized plasmas}
%==============================================================================

We introduce here the collision frequencies between charged and uncharged particles  $\nu_{jn}=n_{n0}\sigma_{jn}
v_{{\sss T}j}$ for $j= e, i$,  and the formulas (Mitchner and Kruger 1972; Spitzer 1963)  for the
Coulomb collisions between charged plasma particles:
\[
\nu_{ee}+ \nu_{ei}\! \simeq\! 2 \nei\!  =\! \left[4 n_{e0} (2 \pi/m_e)^{1/2} [e
e_i/(4 \pi \varepsilon_0)]^2 L_{ei}/[3 (\kappa T_e)^{3/2}]\right],
\]
\[
  \nu_{ii}= \left[4 n_{i0} (\pi/m_i)^{1/2} [e_i^2/(4 \pi
\varepsilon_0)]^2 L_{ii}/[3 (\kappa T_i)^{3/2}]\right].
\]
All frequencies will be given  in Hz  and $L_{ei}=\log [12 \pi \varepsilon_0(\varepsilon_0/n_{i0})^{1/2} (\kappa
T_e)^{3/2}/(e e_i^2)]$ describes the Coulomb logarithm.

Several comments are noteworthy before continuing the derivation. Using the full quantum theory as well as the
semi-classical approach, the elastic proton-hydrogen ($H^++H$) collision cross section $\sigma_{in}$ is calculated by Krstic and
Schultz (1999), and its integral value at $0.5\;$eV is about $1.8 \cdot 10^{-18}\;$m$^2$ for the elastic scattering,
and about $ 10^{-18}\;$m$^2$ for the momentum transfer. As for the electron-hydrogen ($e^-+H$) collisions, the
collision cross section $\sigma_{en}$ is also temperature dependent and the corresponding values can be found in the works of
Bedersen and Kieffer (1971), and  Zecca et al.\ (1996). At energies of $0.5\;$eV it is about $3.5 \cdot
10^{-19}\;$m$^2$,  so that for the elastic scattering we have $\sigma_{in}/\sigma_{en}\simeq 6$.

On the other hand, here we do not include the inelastic collisions that take place in a partially ionized plasma,
like in the photosphere. It can be shown (Vranjes and Poedts 2006) that, in the photosphere, all ions in a unit volume are
recombined many times per second. The three-body recombination (the process of the type $H^+ + e^- + e^-\rightarrow H
+ e^-$) is dominant in this region. At the altitude of $h=500\;$km, the radiative recombination (the process
described by $H^+ + e^-\rightarrow H + h \nu$) and the three-body recombination are of the same order. At higher
altitudes, the radiative recombination becomes the leading loss effect. At $h=1000\;$km, it is by a factor 100 larger
than the three-body recombination.

In addition, the charge exchange between the ionized and neutral hydrogen is frequent. The cross section  (Krstic and Schultz 1999) for the
proton-hydrogen charge exchange $\sigma_{ex}$ at the above given temperatures is about  $5.6 \cdot 10^{-19}\;$m$^2$,
i.e., for hydrogen it is a large fraction ($\simeq 0.3$) of the realistic elastic scattering cross section
$\sigma_{in}$ given above. Note, however, that for some other gases, like He, Ne, and Ar, we have $\sigma_{ex}>
\sigma_{in}$ (Raizer 1991), i.e., the charge exchange cross section exceeds the one for the elastic scattering.
Consequently, due to the inelastic collisions and the charge exchange, neutrals/ions in the plasma spend a part of
their time in the ionized/neutral state, respectively. As a result, the effective collision frequencies are expected
to be even higher than the values that we shall use.

Using the data for a quiet Sun model (Vernazza et al. 1981), in Table~1 we summarize the
values for the electron and proton elastic scattering collision frequencies at
two  altitudes (viz.\ $h=0\;$km, and $h=250\;$km) in the solar photosphere (see
also Vranjes et al.\ 2007). Here, we have taken $B_0=10^{-2}\;$T, the
corresponding temperatures are respectively $T=6420\;$K and $T=4780\;$K, the
electron number densities are $n_0=6.4 \cdot 10^{19}\;$m$^{-3}$ and $n_0=2.7 \cdot
10^{18}\;$m$^{-3}$, and the atomic hydrogen number densities are $n_{n0}=1.17 \cdot
10^{23}\;$m$^{-3}$ and  $n_{n0}=2.3 \cdot 10^{22}\;$m$^{-3}$. We assume that  the proton
and electron number densities are equal.  It is seen that both protons and
electrons are un-magnetized. Note that in Table~1 the collision frequencies
between the plasma species and neutrals are dominant for both electrons and
ions, compared to the frequencies for Coulomb collisions between charged
particles.

It is believed (Priest 1987; Sen and White 1972) that, due to the low temperature, the ions
in the lower photosphere are in fact mainly  metal ions. Sen $\&$  White (1972)
have assumed that the mean mass of these metal ions is 35~a.u.  In that case,
due to the rather different masses of (metal) ions and neutral (hydrogen)
atoms, in calculating the collision frequency it is appropriate to use a more
accurate formula $\nu_{mn}=n_{n0}\sigma_{mn} [m_n/(m_m + m_n)] [8 \kappa
T_m/(\pi \mu)]^{1/2}$, where the index $m$ denotes the metal ion, $n$ denotes
the neutrals (hydrogen), and $\mu=m_m m_n/(m_m+ m_n)$ is the reduced mass. The
calculations may be inaccurate because the collision cross section
$\sigma_{mn}$ is not known. As a guess, we take it as the value for protons
multiplied by $m_m/m_p$.  Taking the layer $h=250\;$km, we find
$\nu_{mm}=6.4\cdot 10^5\;$Hz, $\nu_{mn} = 4\cdot10^8\;$Hz, and
$\Omega_m=2.7\cdot 10^4\;$Hz. Comparing to protons from Table~1, the metal ions
appear to be even less magnetized, i.e., $\Omega_m/\nu_m = 6.6 \cdot 10^{-5}$,
where $\nu_m=\nu_{mm}+ \nu_{mn}$. At $h=0\;$km, we have $\nu_{mm}=1.2\cdot
10^7\;$Hz, $\nu_{mn} = 2 \cdot 10^9\;$Hz, and $\Omega_m/\nu_m\simeq 1.3\cdot
10^{-5}$. The mentioned uncertainty in determining $\sigma_{mn}$ will clearly
not substantially change the fact that the ions are un-magnetized.

%==============================================================================
\section{Physical picture of Alfv\'en waves in a weakly ionized plasma}
%==============================================================================

Following standard textbooks (e.g.\ Chen 1988), in the case of the shear Alfv\'en wave with $\vec B_0= B_0\vec e_z$,
both ion and electron fluids oscillate in the direction of the perturbed magnetic field vector $\vec B_1=B_1 \vec
e_y$. This is due to the $\vec E_1 \times \vec B_0$ drift, which separates neither charges nor masses, and
the direction of the electric field is determined by the Faraday law. The wave is in fact sustained by the additional
polarization drift $\vec v_{pj}= (m_j/q_jB_0^2) \partial \vec E_1/\partial t$ and the consequent Lorentz force $j_x
\vec e_x \times \vec B_0$, which is again in the $y$-direction and has a proper phase shift. Note that the
polarization drift appears as a higher order term due to $|\partial/\partial t|\ll \Omega_i$. It introduces the ion
inertia effects and if it is neglected, then the Alfv\'en wave vanishes. The $\vec E\times \vec B$ term essentially
describes the magnetic field frozen-in property of the plasma. The mode is fully described by the  wave equation
\be
\nabla\times \nabla\times  \vec E_1= \frac{\omega^2}{c^2} \vec E_1 + \frac{i
\omega}{\varepsilon_0 c^2} \vec j_1, \label{we} \ee
the momentum equations for ions and electrons
\[
m_in_i \left[\frac{\partial \vec v_i}{\partial t} + (\vec v_i\cdot\nabla)\vec
v_i\right] = e n_i\left(\vec E + \vec v_i\times \vec B\right) - m_i n_i
\nu_{in} (\vec v_i - \vec v_n)
\]
\be
 - m_i n_i \nu_{ie} (\vec v_i - \vec v_e),\label{e1a}
 \ee
\[
m_en_e \left[\frac{\partial \vec v_e}{\partial t} + (\vec v_e\cdot\nabla)\vec
v_e\right] = -e n_e\left(\vec E + \vec v_e\times \vec B\right)
\]
\be
- m_e n_e \nu_{en} (\vec v_e - \vec v_n) - m_e n_e \nu_{ei} (\vec v_e - \vec
v_i),\label{e1b}
 \ee
and the corresponding equation for neutrals
 \be
m_nn_n \left[\frac{\partial \vec v_n}{\partial t} + (\vec v_n\cdot\nabla)\vec
v_n\right] = -m_n n_n\nu_{ni}(\vec v_n-\vec v_i) -m_n n_n\nu_{ne}(\vec v_n-\vec
v_e).
 \label{e1c}
\ee

Usually, the viscosity may be omitted, and this is valid even for the neutrals.
We note that the dynamic viscosity coefficient for the atomic hydrogen $\mu$
for the layers $h=0$ and $250\;$km (with the temperatures $T_0=6420\;$K and
$T_0=4780\;$K), can be taken (Vargaftik et al. 1996) as $6.5 \cdot 10^{-5}\;$Ns/m$^2$ and
$5.2 \cdot 10^{-5}\;$Ns/m$^2$, respectively. The corresponding kinematic
viscosity coefficient $\eta=\mu/(m_nn_n)$ is 1.34 m$^2$/s. The ratio $\eta
k^2/\nu_{ni}$ for the wave-lengths of interest is very small and the viscosity
effects appear negligible in spite of such a low ionization. Further, using the
momentum conservation in the friction force terms yields $\nu_{ni}=m_i n_0
\nu_{in}/(m_n n_{n0})$, and at $h=250\;$km and  for $m_i=m_n$ we obtain
$\nu_{ni}=3\cdot 10^4\;$Hz. The thermal effects may also be omitted, as will be
explained below.

In the absence of collisions, the response of a plasma to the magnetic and electric field perturbations is
instantaneous, and a volume element of the plasma moves in the previously described manner. In such an ideal case,
the energy flux of the Alfv\'en wave is given by
\be
F_{id}=m_i n_0 v_i^2 c_a/2. \label{f} \ee
Here, $v_i$ is the leading order $\vec E\times \vec B$ perturbed ion velocity.
Its amplitude is given by $v_i=E_1/B_0$. Using the Faraday law we have
$E_1=\omega B_1/k$, hence $v_i=c_a B_1/B_0$. For the estimate only, we assume
small  perturbations of the magnetic field, viz.\ around $1 \%$ (a comment
on larger perturbations will be given later on). For the parameters at
$h=250\;$km, this yields $c_a=B_0/(\mu_0 n_{i0} m_i)=1.3\cdot 10^5\;$m/s.
Consequently, the perturbed plasma (ion) velocity is $v_i=10^{-2} c_a=1.3\cdot
10^3\;$m/s. The wave energy flux in the ideal case, and for $m_i=m_p$, becomes
$F_{id}=5.3\cdot 10^2$ J/(m$^2$s). Setting $m_i=35 m_p$ yields  $F_{id} \simeq
90$ J/(s m$^2$).

\begin{figure}
%  \includegraphics[bb=0 0 370 226,width=0.8\columnwidth,clip=]{magn-2.eps}
%  \includegraphics[bb=409 0 775 226,width=0.8\columnwidth,clip=]{UNMAG.eps}
\includegraphics[bb=0 0 370 222,width=0.8\columnwidth,clip=]{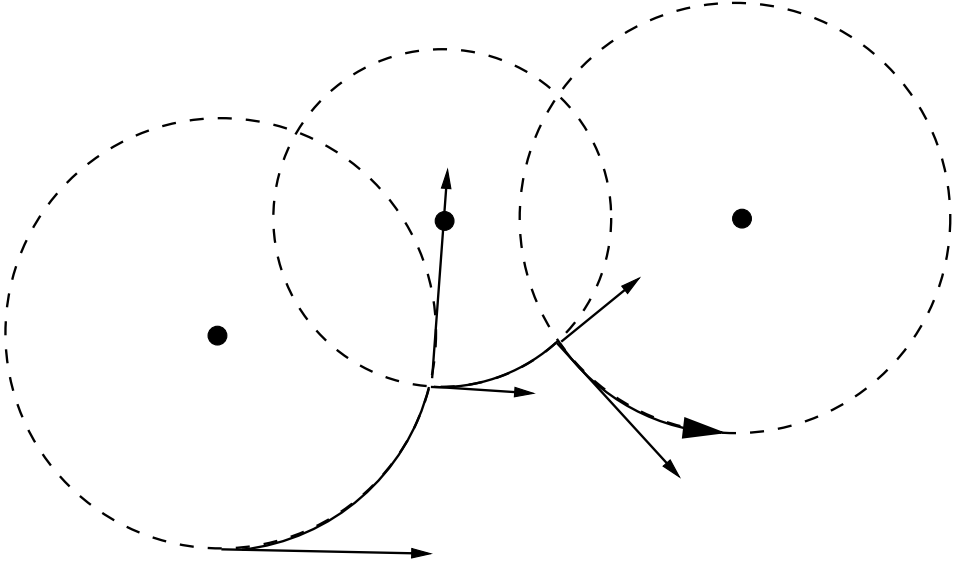}
  \caption{Schematic presentation of the motion of a charged particle  in  non-magnetized  plasma.
  }
  \label{fig:sketch}
\end{figure}

Collisions may heavily alter the motion of the perturbed electron and ion fluids. The plasma response to the
electromagnetic (Alfv\'en-type) perturbations in fully and weakly ionized plasmas is essentially different from
various points of view.  Here, we present some facts that  should help in understanding the physics involved in the
description of the Alfv\'en waves in the partially ionized plasmas, and in particular in the photosphere.
\begin{enumerate}
\item From Eq.~(\ref{e1a}) we see that the ratio of the Lorentz and the friction forces (for predominant
      ion-neutral collisions and in the case of initially unperturbed neutrals), is $\Omega_i/\nu_{in}$.
      For the given photospheric plasma this is $\sim 1/10^3$.
\item The  motion of an un-magnetized charged particle is depicted in Fig.~1.  Arrows denote the tangential direction at the moment of collision when the particle  switches to another  gyro-orbit   with a possibly different velocity (indicated by different gyro-radius).  A collision occurs after a very tiny
  fraction (largely exaggerated here) of the gyro orbit has been traveled. According to
      the numbers from the Table~1, the particle trajectory along a gyro-orbit around one specific magnetic line is only about $1/10^3$ part of the full circle.  Hence, in the given case the path of
      the particle between two collisions is nearly a straight line, like in the case when the magnetic field is absent. In
      fact, the particle never makes a full rotation. The motion is similar in the fully ionized plasmas, however, there it
      is related to ion-ion collisions (i.e., viscosity, not to friction).
\item Contrary to the case of fully ionized plasma where all particles in a volume element move together due to the
      given electric field while still colliding with each other, in the present case the collisions introduce a
      limitation. In the beginning only charged particles are supposed to move due to the applied electric field, while
      neutrals have  a tendency to  stay behind. If $\nu_{in}\gg \Omega_i, \omega$, each plasma particle collides many
      times within the theoretical gyro-rotation, or within the assumed wave oscillation.
\item Contrary to the viscosity, which is of primary importance for short scale processes, the friction is more
      effective in the opposite limit, i.e., at smaller wave-numbers (and also at larger wave-periods) an ion is subject to
      larger number of collisions with neutrals within one oscillation period.
\end{enumerate}

Below, we discuss the effects of collisions
      in detail. According to the presently widely accepted physical description of Alfv\'en waves in partially ionized plasmas
(Kulsrud and Pierce 1969; Haerendel 1992; De Pontieu and  Haerendel 1998;   P\'{e}cseli and   Engvold 2000), the effects of neutrals are the following: a)~for a relatively small amount of neutrals (or for high frequency short wavelengths), the damping of
the mode is proportional to the collision frequency $\nu_{in}$, more collisions
increases the friction (Kulsrud and  Pierce 1969); and b)~in a very weakly ionized plasma the
collisions are numerous and the whole fluid moves together. In this case, the
stronger the collisions the better locking of the gas-plasma fluid, and the damping
of the wave (which is now proportional to $1/\nu_{in}$) vanishes. The Alfv\'en
velocity in such a mixture includes the total fluid density $m_i n_i + m_n
n_n$. The dispersion equation of the Alfv\'en wave is, according to De Pontieu
$\&$  Haerendel (1998):
\[
\frac{\omega}{k}=c_{\sss A}\left(1- i \frac{m_n n_n}{m_n n_n + m_i n_i}
\frac{\omega}{\nu_{ni}}\right)^{1/2}\!\!\!, \!\!\quad
 c_{\sss A}=\frac{B_0}{[\mu_0 (m_i n_i + m_n n_n)]^{1/2}}.
\]

Taking these statements as facts, here we suggest the following item to be added in order to complete the physical
picture, thus c)~the perturbed velocity of the gas-plasma mixture may be drastically reduced in a weakly ionized plasma
(like the photosphere), and, consequently, {\em the wave energy flux becomes very small}.

\begin{figure} [!ht]
\includegraphics [width=8.cm] {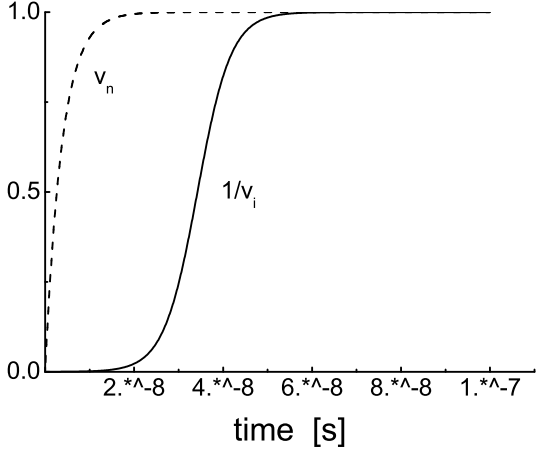}
  \caption{ Relaxation of the ion and neutral velocities
  (normalized to $v_c=(\nu_{in} V_n + \nu_{ni} V_i)/(\nu_{in} + \nu_{ni})$) due to
  collisions, for parameters appropriate for the
  solar photosphere. The initial velocities for neutrals and ions are
  respectively 0 and $1.3 \cdot 10^3\;$m/s.}
\end{figure}

To describe what happens in reality, we here assume {\em the same magnitude of
the magnetic field perturbation as in the ideal case discussed above, i.e.,
taking it as  $ 1 \%$}. Due to the perturbed magnetic field, there appears the
electric field as described above, and the consequent ion motion in the same
direction as the perturbed vector $\vec B_1$. We denote this perturbed $\vec E\times \vec
B$-drift velocity of ions  by $V_i$. Note that this velocity is the
same for electrons, and that we are speaking about fluid velocities.  The
relaxation velocity of neutrals and ions can be obtained from the following
estimates (Milic 1970). Assume that in the starting moment the unit volume of
the neutrals have a different velocity $V_n$. In view of the huge difference in
mass, we neglect electrons for simplicity. The collision frequency (see Table~1)
is extraordinarily high, of the order of $\sim 10^{9}$ Hz. Compare this with the
theoretical gyro-frequency for ions $\Omega_i \sim 10^{6}$ Hz. Knowing that the
wave frequency must be much smaller, the frequency ordering that  we have here
is:
\be
\omega\ll \Omega_i\ll \nu_i. \label{or}
\ee
As a result, we can take the starting/maximal value of the ion velocity $V_i$
and estimate for the value it will take within the collisional time.

 In view of the ordering (\ref{or}), the time dependence of the velocities of the two fluids (ions and neutral) in relative motion,  after the initial movement of the plasma due to electromagnetic perturbations has taken place (regardless of the origin of these perturbations),  is determined  mainly  by the friction, and  can be obtained from the following equations:
\be
\partial \vec v_n/\partial t= \nu_{ni} (\vec v_i - \vec v_n), \quad
\partial \vec v_i/\partial t= \nu_{in} (\vec v_n - \vec v_i). \label{par}
\ee
Simple combinations of these equations yield two integrals of motion:
\[
\nu_{in} \vec v_n + \nu_{ni} \vec v_i\!=\! \nu_{in} \vec V_n + \nu_{ni} \vec
V_i, \quad \! \vec v_n-\vec v_i\!=\! \left(V_n- \vec V_i\right) \exp[-(\nu_{in}
+ \nu_{ni})t].
\]
This further  yields
 \be
\vec v_n=\frac{\nu_{in} \vec V_n + \nu_{ni} \vec V_i}{\nu_{in}+ \nu_{ni}} +
\frac{\left(\vec V_n - \vec V_i\right) \nu_{ni}}{\nu_{in}+ \nu_{ni}}
\cdot\exp[-(\nu_{in} + \nu_{ni})t], \label{pari} \ee
\be
 \vec v_i=\frac{\nu_{in} \vec V_n + \nu_{ni} \vec V_i}{\nu_{in}+ \nu_{ni}} -
\frac{\left(\vec V_n - \vec V_i\right) \nu_{in}}{\nu_{in}+ \nu_{ni}}
\cdot\exp[-(\nu_{in} + \nu_{ni})t]. \label{parn} \ee
It is seen that the two velocities relax very quickly towards the first term on
the right-hand sides in these expressions. Taking $m_i\sim m_n$ and  $\vec
V_n=0$, we obtain for the relaxed (common) velocity for both species
 \be
v_c = V_i\frac{ \nu_{ni}}{\nu_{in} + \nu_{ni}} = V_i \frac{n_{i0}}{n_{i0}+
n_{n0}}\simeq V_i \frac{n_{i0}}{n_{n0}} .\label{com} \ee
Hence, if ions start to  move due to the electromagnetic force caused by the wave, the strong friction will result in
a common  velocity, which  i)~is achieved very quickly, and ii)~is much below the starting velocity of the ion
fluid. This behavior is presented in Fig.~2 for the photospheric parameters at $h=250\;$km.

\begin{table*}
\caption{Parameters of waves (wave-lengths $\lambda$ in km and frequencies in Hz) propagating through the chromosphere for two
different altitudes $h$ (in km).} \label{table:2} \centering
\begin{tabular}{c c c c     }     % 6 columns
\hline\hline
 $h=1065 $  \\
 \hline
 $\lambda$  & $\omega$ & $ k c_a$ & $\omega_i/\omega_r$  \\
\hline
   0.1   & $311 -1222 i$ &$44855$     & $ 3.9$  \\
   1 & $327 -45 i$ & $4485$  & $0.14$  \\
      10   & $33.1 -0.45 i$ &$448.5$     & $ 0.014$  \\
        100   & $3.3 -0.0045 i$ &$44.85$     & $ 0.0014$  \\
          500   & $0.66 -0.0002 i$ &$8.97$     & $ 0.0003$  \\
 \hline
\end{tabular}
\hspace{0.3cm}
\begin{tabular}{c c c c     }     % 6 columns
\hline\hline
 $h=1990 $   \\
 \hline
 $\lambda$  & $\omega$ & $ k c_a$ & $\omega_i/\omega_r$  \\
\hline
   0.1   & $69666 -732 i$ &$69829$     & $ 0.01$  \\
   1 & $6891 -722 i$ & $6983$  & $0.1$  \\
      10   & $371 -94.5 i$ &$698.3$     & $ 0.25$  \\
        100   & $36.4 -0.9 i$ &$69.8$     & $ 0.025$  \\
          500   & $7.3 -0.04 i$ &$14$     & $ 0.005$  \\
 \hline
\end{tabular}
\end{table*}

Here, for the same perturbation of the magnetic field ($1 \%$), we have
$v_c=1.15\cdot 10^{-4} V_i=0.15\;$m/s. Compare this to the velocity in the
ideal case $V_i=1.3 \cdot10^{3}\;$m/s. Note also that both $V_i$ and $v_c$ are
below/much below the sound velocity $c_s=8.9 \cdot 10^3\;$m/s, respectively.
Hence, neglecting the pressure (compressibility) effects is justified. Because
$v_c$ is so small, in Fig.~2 we normalize velocities to $v_c$ and give the plot
for $v_n$ and  $1/v_i$. It is seen that  the velocities of both neutrals and
ions relax towards the same (normalized) value ($=1$) within a time interval
that is many orders of magnitude shorter than the wave oscillation period. As
a result, using (\ref{f}),  we have the flux in the weakly ionized plasma (for
$m_i=m_p$) given by
\be
F=\frac{1}{2} (m_i n_i + m_n n_n) c_{\sss A} v_c^2=F_{id} \left(\frac{m_i
n_i}{m_i n_i+ m_n n_n}\right)^{3/2}. \label{rf} \ee
For the given parameters in the photosphere this gives \be
 F\simeq 10^{-6}\cdot
F_{id}=5.3 \cdot  10^{-4} \,\,J/( m^2s). \label{rf2} \ee
It is seen that the actual flux is always small for any realistic amplitude of
perturbations. For example, even taking exceptionally strong magnetic field
perturbations, e.g.\ $B_1=B_0$, yields  $F\simeq 5$ J/(m$^2$s). Consequently,
regardless of the physical mechanism for eventual excitation of the Alfv\'en
waves in the photosphere, the expected amplitude of the perturbed velocity is
of the order of $0.1\;$m/s, and the energy flux of the waves is about one
million times smaller than the one obtained from the ideal  models that assume a perfect coupling between the plasma and magnetic field (i.e., ignoring  the effects of collisions  between ions and neutrals, and the consequent  weak magnetization of plasma species).

The estimated flux presented above is obtained for $m_i=m_p$. Taking the more realistic value $m_i=35 m_p$ \cite{sw},
we obtain only $F=0.02$ J/(m$^2$s). Assuming in addition a stronger magnetic perturbation of $10 \%$, we obtain
$F=2$ J/(m$^2$s) and the common velocity amplitude $v_c\simeq 9\;$m/s. The actual flux may have larger values, e.g.,
due to stronger magnetic field perturbations, but the linear wave theory then becomes unapplicable.

Since  the electromagnetic force still acts on the plasma volume in the time interval $ \nu_{in}^-1$, after the initial movement of the plasma has taken place, one could claim that the flux may be higher. Yet, in view of the frequency ordering (\ref{or}), which implies a difference of many orders of magnitude, the inclusion of this  additional electromagnetic  effect in Eq. (\ref{par}) is insignificant. In fact, it is questionable and indeed unlikely that the ions can really achieve  the assumed starting perturbed velocity  $\vec V_i$ in the first place. This is because  the assumed value for $\vec V_i$ follows from the ideal case discussed above,  with time and spatial scales determined by $\omega^{-1}$ and $ k^{-1}$, respectively, resulting in the characteristic velocity $c_a$, while in the collisional case that  we have here, these scales are determined by $\nu_{in}^{-1}$ and $\kappa_f^{-1}$, where $\kappa_f^{-1}=v_{\sss{T}i}/\nu_{in}$ is the ion mean free path. Hence, the characteristic velocity that  now appears instead of $c_a$ is $v_{\sss{T}i}=(\kappa T_i/m_i)^{1/2}$  and it  is  about 2 orders of magnitude lower than $c_a$, and a realistic  flux should be even smaller than the value obtained earlier.   One could also  argue that the case discussed above, $\vec V_{n}=0$,   may look the least favorable for the propagation of the wave because neutrals are initially usually  in the state of motion. Clearly this does not change anything, because in this case, due to the strong collisions,  the ions will nearly be in the  same state of motion (see in the Sect. 4 below), while the ion velocity   $\vec V_i$ would still describe  an access ion momentum  obtained due to the electromagnetic perturbation, which neutrals initially do not take part in.

%==============================================================================
\section{Discussion}
%==============================================================================

Standard estimates of the wave energy flux through the solar photosphere assume a plasma velocity in the photosphere
of the order of $1\;$km/s. This implies two effects:  that plasma particles move with the observed speed of the
convective motion, and  that the motion of plasma species involves the magnetic field perturbations due to
frozen-in magnetic field effect. The first effect is only partly satisfied. If in the equilibrium  neutrals move
perpendicular to the magnetic field, say in the $x$-direction, the plasma particles will move also due to the
friction effect. The induced velocities of ions and electrons can be calculated from Eqs.~(\ref{e1a}) and (\ref{e1b})
reading
\be
\vec v_{i\bot 0} = \alpha_i\left( -\frac{\nin}{\Omega_i} \vec e_z \times \vec
v_{n\bot 0} + \frac{\nin^2}{\Omega_i^2} \vec v_{n\bot 0}\right), \label{e3} \ee
and
\[
\vec v_{e\bot 0} = \alpha_e\left( \frac{\nen}{\Omega_e} \vec e_z \times \vec
v_{n\bot 0} + \frac{\nei}{\Omega_e} \vec e_z\times \vec v_{i\bot 0} +
\frac{\nen \nu_e}{\Omega_e^2} \vec v_{n\bot 0} \right.
\]
\be
\left.
 + \frac{\nei
\nu_e}{\Omega_e^2} \vec v_{i\bot 0}\right), \label{e4} \ee
where
\[
\alpha_{e, i}= \frac{1}{1+ \nu_{e, in}^2/\Omega_{e, i}^2}, \quad \nu_e= \nue.
\]
The ion drag velocity (in the $x$-direction)  and the drift component (in the
$y$-direction) become, respectively,
\be
v_{i0, drag}=v_{ix0}=\frac{1}{1 + \Omega_i^2/\nin^2} v_{nx0}, \label{dv} \ee
and \be
v_{i0, drift}=v_{iy0}= -\frac{\nin}{\Omega_i} \frac{v_{nx0}}{1+
\nin^2/\Omega_i^2}=-\frac{\Omega_i}{\nin} v_{i0, drag}.
 \label{dv2} \ee
The corresponding electron components are
\be
v_{e0, drag}\!=\! v_{ex0}\!=\!\alpha_e v_{nx0}\frac{\nu_e}{\Omega_e}\left[
\frac{\nu_{en}}{\Omega_e} + \frac{\nu_{ei}}{\Omega_e} \left(1 +
\frac{\Omega_e\Omega_i}{\nu_e \nu_{in}}\right) \left(1+
\frac{\Omega_i^2}{\nu_{in}^2}\right)^{-1}\right], \label{dvex} \ee
and
\be
v_{e0, drift}=v_{ey0}= \alpha_e v_{nx0}\frac{\nu_{en}}{\Omega_e}\left[ 1+
\frac{\nu_{ei}}{\nu_{en}}\left(1 - \frac{\Omega_i\nu_e}{\Omega_e
\nu_{in}}\right) \left(1+ \frac{\Omega_i^2}{\nu_{in}^2}\right)^{-1}\right].
\label{dvey} \ee
The induced ion and electron velocities are not necessarily equal, implying the
presence of equilibrium currents. For the same parameters used in Table~1 and
taking the  neutral velocity of $500\;$m/s, at $h=250\;$km we have the drag and
drift velocities for electrons 315 and $240\;$m/s, respectively. The ion drag
velocity is almost equal to the neutral velocity. This is all  due to the fact
that the plasma particles are un-magnetized, $\Omega_i/\nu_i=3.6 \cdot
10^{-3}$, $\Omega_e/\nu_e=0.76$. However, due to the same reason the frozen-in
condition is far from reality and the ion/electron motion perpendicular to the
magnetic lines does not necessarily involve the appropriate movement of the
magnetic lines. The actual motions develops as described in the previous
section.

In view of the item b) discussed in Sect. 3 \cite{kupi}, such an upwards
propagating wave is very weakly damped in the photosphere (the damping is
proportional to $1/\nu_{in})$. This holds provided that the wavelengths exceed
a certain minimal value. However, it will in fact be more strongly damped in
the upper layers, e.g., in the chromosphere where the amount of neutrals
decreases but the damping is proportional to $\nu_{in}$. For the chromosphere
this can be directly demonstrated by solving the dispersion equation that
follows from (\ref{we}) where the perpendicular currents are calculated from
Eqs.~(\ref{e1a})-(\ref{e1c}). The expressions are very lengthy and we shall not
give them here.

As an example, assuming the wave propagating towards the chromosphere, the dispersion equation is solved for several
wavelengths $\lambda$, with all collision frequencies included, at the altitude $h=1065\;$km where  \cite{ver}
$T=6040\;$K, $n_{n0}=1.71\cdot  10^{19}\;$/m$^3$, and $n_{0}= 9.35\cdot 10^{16}\;$/m$^3$, and at the altitude
$h=1990\;$km where  $T=7160\;$K, $n_{n0}= 10^{17}\;$/m$^3$, and $n_{0}= 3.9\cdot 10^{16}\;$/m$^3$. The results are given
in Table~2. It is seen that shorter  wavelengths are more damped at lower altitudes. In the same time, longer
wavelengths (i.e., those that are presumably better transmitted by the photosphere)  are in fact more damped at
higher altitudes. This mode behavior is in agreement with the model of Kulsrud $\&$ Pierce (1969). However, this trend
certainly can not continue because neutrals vanish at still higher altitudes.

We stress that the equilibrium parameters change with the altitude and for the
large wavelengths the model becomes violated. A numerical approach should give
more reliable results. Such an approach could help explain how and where the
Alfv\'en waves, that were recently detected  in the chromosphere \cite{dp2},
are generated.

Our analysis is based on the presence of a temperature minimum in which most of
the plasma is neutral, which is predicted by hydrostatic models  averaged
in space and time, such as VAL and FAL. If flux tubes for some reason lack
this temperature minimum, the analysis we present here may not be an accurate
description of how Alfven waves are generated in the photosphere.

%==============================================================================
\section{Conclusions}
%==============================================================================

The physics of a multi-component weakly ionized plasma, like the one in the
solar photosphere, is highly complex. Various aspects of this complexity have
been pointed out in Sects. 2-4. For a temperature of about $0.5\;$eV, typical
for the photosphere, there is a plethora of processes that take place and that
are nontrivial to include in an analytical work like the one presented here.
Among others, these include the elastic and inelastic collisions, the charge
exchange being an important sort of the latter, which imply the creation and
loss of plasma particles. Yet, in spite of that, some conclusions regarding the
importance  of the electromagnetic Alfv\'en-type perturbations in such weakly
ionized environments can be made with some certainty. The important conclusion
is that if we assume   Alfv\'en-type waves  generated around  the temperature
minimum, in fact their amplitudes are such that the wave energy flux is very
small. The main reason for this is  ion collisions, which are so frequent that
ions almost do not feel the effects of the magnetic field. As seen from Fig.~1,
in such an environment the ion motion is very similar to the Brownian motion of
atoms and molecules in a gas.  The physics presented here should be taken into
account in the estimates of the role of the Alfv\'en waves generated in the
solar photosphere in coronal heating scenarios. However, the solar photosphere
is only a thin plasma layer and the parameters in the solar atmosphere change
with the altitude and so does the physics of the Alfv\'en waves.
%As a matter of
%fact, this is already so in the lower chromosphere where strong Alfv\'en-type
%perturbations have very recently been observed \cite{dp2}.
Our analysis suggests that if these waves are generated below the chromosphere, they cannot
probably  be generated around the temperature minimum, but perhaps  would
have to come from lower down, i.e., below the surface where the plasma is again
much more ionized and the ion-neutral collisions are not significant.

Acknowledgements: 

These results are  obtained in the framework of the projects G.0304.07 (FWO-Vlaanderen), C~90203 (Prodex),
GOA/2004/01 (K.U.Leuven),  and the Interuniversity Attraction Poles Programme - Belgian State - Belgian Science
Policy.

%==============================================================================

%==============================================================================


\begin{thebibliography}{}
\bibitem{bed} Bedersen, B., $\&$  Kieffer, L. J. 1971, Rev. Mod. Phys.,  43, 601
\bibitem{chen} Chen, F. F. 1988,   Introduction to Plasma
Physics and Controlled Fusion (Plenum Press, New York), pp. 136-142
\bibitem{dp}  De Pontieu, B., $\&$  Haerendel, G. 1998,  A$\&$A, 338, 729
%\bibitem[De Pontieu et~al. 2007]{dp2}  De Pontieu, B., McIntosh, S. W., $\&$ Carlsson, M. et~al. 2007,  to be published
\bibitem{haer} Haerendel, G. 1992, Nature,  360, 241
\bibitem{hol2} Hollweg, J. V. 1981, Sol. Phys.,  70, 25.
 \bibitem{js} Jephcott, D. F., $\&$  Stocker, J. 1962,
J. Fluid Mech., 13, 587
\bibitem{krs}  Krstic, P. S., $\&$   Schultz, D. R. 1999,   J. Phys. B: At. Mol. Opt. Phys., 32, 3485
\bibitem{kupi}  Kulsrud, R., $\&$   Pierce, W. P. 1969,  ApJ, 156, 445
\bibitem{mil2} Milic, B. S. 1970,    Statistical
Physics (in serbian)  (Naucna knjiga,  Beograd) pp. 149-152
\bibitem{mk} Mitchner, M., $\&$  Kruger, C. H. 1973,    Partially Ionized Gasses (John Willey and Sons, New York) p. 413
\bibitem{pec}  P\'{e}cseli,  H., $\&$   Engvold, O. 2000,  Sol. Phys., 194, 73
\bibitem{pr}  Priest, E. R. 1987,  Solar
magnetohydrodynamics (D. Reidel Pub. Co., Dordrecht)
\bibitem{pud} Pudritz, R. E. 1990, ApJ,  350, 195
\bibitem{reiz}  Raizer, Y. P. 1991,    Gas discharge
physics (Springer-Verlag, Berlin Heidelberg), p. 25
  \bibitem{sw} Sen, H. K., $\&$   White, M. L. 1972,
Sol. Phys., 23, 146
\bibitem{r3} Spitzer, L. 1962,    Physics of Fully
Ionized Gasses (Interscience Publishers,  New York-London) p. 146
 \bibitem{tan} Tanenbaum, B. S., $\&$   Mintzer, D. 1962,
 Phys. Fluids, 5, 10
\bibitem{var}  Vargaftik, N. B., Vinogradov, Y. K., $\&$  Yargin, V. S. 1996,   Handbook of Physical Properties of Liquids and Gases (Begel House, New York-Wallingford) p. 59
 \bibitem{ver} Vernazza, J. E., Avrett, E. H.,
$\&$  Loeser, R. 1981, ApJS, 45,  635
\bibitem{vr1} Vranjes, J., $\&$   Poedts, S. 2006, Phys. Lett. A, 348, 346
\bibitem{vprl} Vranjes, J., Poedts, S., $\&$ Pandey, B. P.
2007,  Phys. Rev. Lett.  98, 049501
 \bibitem{watts} Watts, C., $\&$  Hanna, J.  2004,
Phys. Plasmas,  11, 1358
\bibitem{woods} Woods, L. C. 1962,
J. Fluid Mech., 13, 570
\bibitem {zec}  Zecca, A.,   Karwasz, G. P., $\&$ Brusa, R. S. 1996,   Riv. Nuovo Cim.,  19, 1

\end{thebibliography}
\end{document}